# Language Preferences and Practices in Multilingual EdTech: Flexible Primary Language Use with Secondary Language Support

Christine Kwon*, Phenyo Phemelo Moletsane*, Michael W. Asher, Dieyu Ouyang, Lingkan Wang, Debbie Eleene Conejo, John Stamper, Paulo F. Carvalho, and Amy Ogan
ckwon2@andrew.cmu.edu, pmoletsa@andrew.cmu.edu, masher@andrew.cmu.edu, dieyuo@alumni.cmu.edu, lingkanw@andrew.cmu.edu, dconejo@andrew.cmu.edu, jstamper@cmu.edu, pcarvalh@andrew.cmu.edu, aeo@andrew.cmu.edu
Carnegie Mellon University
*These authors contributed equally to this work

**Abstract:** The benefits of learning in one's mother tongue are well documented, yet colonial languages dominate education, marginalizing local languages and limiting access for learners who rely on their mother tongue for understanding. With the rapid growth of educational technology, there is potential to integrate multilingual instruction supporting both colonial and local languages. This study is part of a larger quasi-experiment conducted in Uganda, where learners could choose to learn in English, Leb-Lango (a local language), or in Hybrid mode (a combination of both) in a remote EdTech course. We examined how learners who chose the Hybrid option navigated English and Leb-Lango. While many Hybrid learners did not consistently use both languages, those who did persisted longer in the course. Learners also shared how they managed language complexities. We provide the first empirical evidence of learner agency in bilingual remote EdTech instruction and offer insights for designing inclusive multilingual learning solutions.

## Introduction and Related Work

Implementing languages of instruction is a persistent challenge and debate within marginalized learning spaces, especially in post-colonial majority world contexts, where the linguistic landscape is rich with diverse local languages spoken across various communities alongside colonial languages (Wolff, 2017). Colonial influences continue to dominate most learning spaces, while local language practices remain undervalued within education (Skutnabb‑Kangas & McCarty, 2008). This reflects a continuing perspective from prior literature investigating colonial and local languages of instruction in education, associating colonial languages with power and higher privilege (Liddicoat & Heugh, 2014; Ndebele, 2014; Kukulska-Hulme et al., 2023), and local languages with limited use in their homes and communities (Trudell, 2007; Wildsmith-Cromarty et al., 2023). Prior studies investigating the languages of instruction within marginalized African learning spaces have shown that many students and families prefer education led by colonial languages due to their association with prestigious life opportunities (Muthwii, 2004; Trudell, 2007).

However, previous work investigating local languages of instruction has shown that local languages can provide a relevant and meaningful learning experience for marginalized and remote learners (Buhl-Wiggers et al., 2023; Msimanga & Lelliott, 2013; Nomlomo, 2007; Piper et al., 2016). For instance, Piper et al. (2016) conducted a medium-scale randomized controlled trial (RCT) in Kenya involving first- and second-year primary learners who received instruction in their mother tongues (Kikamba or Lubukusu), showing that those who received the mother tongue intervention showed significant improvements in literacy outcomes in oral reading fluency and reading comprehension compared to those who were only taught in English and Kiswahili. Additionally, translanguaging theory emphasizes the importance of multilingual learners to navigate and integrate multiple languages to construct meaningful understanding and enhance learning outcomes. For instance, Nyimbili & Mwanza (2020) conducted a quasi-experiment among first-grade learners in the Lundazi district in Zambia between students who were taught with translanguaging (experimental group) or monolingual practices (control group), in which learners in the experimental group showed a significant increase in mean scores from pre to post assessments, while those in the control group showed a slight reduction. In South Africa, Msimanga & Lelliott (2013) examined small-group discussions of grade 10 chemistry students to explore how learners engaged with science content when using local languages. They found that learners consistently used their home language (predominantly isiZulu) and engaged in frequent translanguaging practices to make sense of scientific concepts, enhancing engagement and collaborative meaning-making, with learners making claims, questioning peers, and negotiating scientific concepts in their home languages.

Thus, multilingual learning strategies incorporating local language practices may benefit multilingual learners in marginalized communities. Yet, reaching these communities at scale requires accessible delivery mechanisms, and this is where educational technology (EdTech) becomes critical. Prior EdTech literature has demonstrated the effectiveness and potential of delivering learning opportunities for remote and marginalized communities by leveraging widely adopted technologies, including interactive radio instruction (IRI) (Verlumun Celestine et al., 2024) and text message-based learning applications (Breazeal et al., 2016; Khan et al., 2019; Kizilcec et al., 2021; Wahyuni et al., 2024). This opens up opportunities for similar EdTech systems to include multilingual instruction with both colonial and local languages to reach multilingual remote communities. However, current EdTech solutions in remote and marginalized regions are mostly conveyed in colonial languages (Omojola, 2009). Prior work investigating EdTech and languages, particularly within marginalized regions, has shown a close connection between English and technology (Kukulska-Hulme et al., 2023; Nkoala, 2024), emphasizing that the dominance of colonial languages of instruction in EdTech can further exclude marginalized learners of local languages (Kukulska-Hulme et al., 2023; Liddicoat & Heugh, 2014).

Nonetheless, some studies have explored comparable approaches to multilingual instruction in EdTech solutions. For instance, Mohammed and Mohan (2013, 2015) developed a localized intelligent tutoring system that taught programming using English-based Creoles, and conducted multiple case studies in Trinidad and Tobago showing that students preferred systems integrating cultural and linguistic context, and valued the ability to control the degree of localization via a "cultural density slider." Similarly, Jantjies & Joy (2012, 2013) introduced bilingual mobile learning tools to high school mathematics and science learners in South African schools, enabling students to access content in both English and local languages and create bilingual audio notes. Findings showed that most learners tended to use both languages to engage with content, reporting that flexible access to learning in both languages helped them articulate and understand concepts more effectively. While these studies offer tentative insights into how learners in neighboring contexts navigate multilingual and cultural inclusion in their learning using different EdTech solutions, there remains a limited understanding of how to design EdTech applications that support multiple languages effectively and meaningfully for these learners (Jantjies & Joy, 2012; Kukulska-Hulme et al., 2023). Additionally, they have largely focused on formal learning environments, which provide limited insight into how learners exercise agency when choosing to learn in both colonial and local languages. Hence, there remains a gap in understanding how to effectively implement and navigate colonial and local languages of instruction in EdTech solutions that address the multilingual needs and contexts of learners who are further marginalized.

Our work extends this to informal learning environments. This study is part of a larger quasi-experiment conducted in a small rural region in Uganda in 2024, in which learners could choose to learn in English, Leb-Lango (a local language), or both Leb-Lango and English (referred to as Hybrid) in a remote course delivered via radio and phone-based technology. In this paper, we focus on learners in the Hybrid condition who opted for both Leb-Lango and English to understand their learning experiences across the two languages. We employed a mixed-methods approach, analyzing course log data and conducting interviews with 29 participants in the Hybrid group. This work is guided by the following research question: **How do learners navigate colonial and local languages of instruction when given the option to learn in both?**

Our findings deconstruct the perceptions and explorations of Hybrid learners using instruction in English and Leb-Lango during the remote course. While our findings initially indicate a significant preference for learning only in English among a plurality of Hybrid learners, a more in-depth investigation revealed that Hybrid learners who explored both English and Leb-Lango demonstrated positive learning outcomes, such as higher course engagement compared to Hybrid learners who did not engage with both languages. Hybrid learners also provided contextual insight into how they considered various language complexities and ways to use English and Leb-Lango instruction during the remote course, including implicit bilingual strategies that employ the use of both languages. The main contribution of our research is to provide empirical evidence of how remote learners choose to navigate and engage with colonial and local languages of instruction in remote EdTech learning environments, providing EdTech developers and instructors with insights to design more inclusive and meaningful EdTech solutions for learners in multilingual marginalized communities.

## Methods

This work is part of a multi-year collaboration with an educational technology company in Northern Uganda. The authors of the paper were not involved in the recruitment, consent, or data collection processes, all of which were conducted by our collaborators in Uganda. The study was approved by a Ugandan ethics committee.

### Context, Study Design and Data Analysis

The data analyzed in this study come from a quasi-experimental study conducted in Lira District, Northern Uganda. Uganda is a multilingual country with over 60 indigenous languages (Criper & Ladefoged, 1971), alongside two official but non-indigenous languages: English and Kiswahili (Schmied & Mesthrie, 2008). The language-in-education policy promotes bilingual instruction where local languages are used as the medium of instruction in the first three years of primary school, while English becomes the medium of instruction from Primary Four onwards through secondary and higher education (Altinyelken, 2010; National Curriculum Development Centre, 2006). The population in the Lira district is predominantly from the Lango ethnic group, and the primary language spoken is Leb-Lango, a minority language used by the participants of this study. Learners were recruited for the remote course through radio advertisements and asked to complete a recruitment questionnaire via text message. They were given the option to consent to both enroll in the course and allow their learning data to be used for course improvement and research purposes. A total of 2,931 learners were enrolled in the remote course. All learners registered for the course using their phones and indicated their preferred instructional language during registration. Based on this choice, they were assigned to one of three groups: English, Leb-Lango, and a Hybrid option, which provided the flexibility to choose between English and Leb-Lango.

We analyzed data from a version of the course delivered between August 5th and November 24th of 2024. The course was delivered through a low-tech platform and designed to provide interactive STEM education to participants in rural Uganda, where access to the internet, smartphones, computers, and academic materials is often limited (Kwon et al., 2023). As such, it relied on two widely available tools in rural Africa to reach remote learners: basic keypad phones and radios (Uganda Bureau of Statistics, 2024). Once registered, learners received alerts through Robocalls and Short Message Service (SMS), which provided information about upcoming content and any required materials. On specific days of the week, learners could tune into a dedicated local radio broadcast, where an instructor delivered lessons and guided learners through the course content. Based on their language choice, learners tuned into the radio broadcast, which was either instructed in English at 12:00 PM East Africa Time (EAT) or in Leb-Lango at 2:00 PM EAT.

The course curriculum followed the steps of the Engineering Design Process, guiding learners to build practical technologies that addressed everyday problems in the community. The course curriculum was divided into a 9 step process: Introduction Step: Basics of STEM education; Step 1: Identify; Step 2: Investigate; Step 3: Brainstorm; Step 4: Plan; Step 5: Create; Step 6: Test; Step 7: Improve; Step 8: Launch. The course we analyzed taught learners how to create a solar-powered food dehydrator, a device that uses solar energy to dry and preserve produce, such as fruits and vegetables. Instructors engaged learners by asking questions via the radio, while learners responded in real time or asynchronously using Unstructured Supplementary Service Data (USSD) menus on their phones or through live call-ins. These phone-based interactions from the USSD were logged, enabling tracking and analytics of learner engagement and performance. Instruction and assessments were delivered in the learner's chosen language option. During the course, learners completed multiple-choice practice questions and multiple-choice weekly assessments based on prior lessons. At the end of the program, learners took a final 16-question multiple-choice assessment delivered via radio and responded using USSD.

To understand how learners navigated both colonial and local languages when given the option to learn in both, we examined their behavior in the Hybrid option of the course. Learners in the Hybrid option could switch languages on their phones through a scheduled language switch feature at the end of each step or within a step by updating the language in the profile menu. For each Hybrid learner, we created a language-use flag from the log data: English only (all responses in English), Leb-Lango only (all responses in Leb-Lango), or Explorers. We define Explorers as learners who used both Leb-Lango and English at least once in their responses. We use the term Explorers because using both languages suggests that learners were testing or trying out the available language options, rather than consistently responding in only one language, and may therefore have been exploring which language to use when responding. We then calculated the proportion of learners in each category relative to all Hybrid-course participants. Next, we examined the trajectories of language switching among Explorers. We aggregated the learners' ordered responses into a sequence of response languages, identified points where the language changed and labeled these positions as start, first, second, and so on up to the maximum number of switches observed. We calculated the frequency of each trajectory across learners as well as the proportion of learners at each position. To further understand how the switching behavior of Explorers evolved across the course, we created a binary indicator for each learner and step showing whether any language switch occurred. We then summarized these indicators across all learners to calculate the proportion of students who switched languages at each step. Finally, we explored how persistence differed among Hybrid learners: those who used both English and Leb-Lango (Explorers) compared to those who used only one language (English only, or Leb-Lango only). For each learner and step, we coded presence as 1 if the learner was present at the step and 0 otherwise. We fitted a generalized linear mixed-effects model to estimate the likelihood of presence at each step, including language behavior (English only, Explorers, Leb-Lango only) and step as fixed effects, with Explorers as the reference category. To account for repeated observations within learners, a random intercept for each learner was included.

To qualitatively understand the perceptions and experiences of Hybrid learners navigating English and Leb-Lango instruction in the remote course, our collaborators and course facilitators conducted interviews with a sample of Hybrid learners (n = 29) after the remote course. Interview participants were selected based on four pre-determined criteria: (1) completed the course and scored above 75% on the final exam, (2) completed the course and scored below 75% on the final exam, (3) stopped course engagement at the midpoint of the course, (4) stopped course engagement before the midpoint of the course. Facilitators conducted and audio-recorded in-person interviews with these learners at their respective homes. Both verbal and written consent were collected from all interviewed participants, including parental consent for learners under the age of 18. The course facilitators conducted semi-structured interviews in the local language, Leb-Lango, then translated all interviews into English and removed all personal identifiers. Interviews averaged about 17 minutes, ranging from approximately 11 to 31 minutes. We conducted a thematic analysis on the post-course interviews, following the thematic framework by Braun & Clarke (2006). Four researchers developed an open codebook using an inductive process, in which codes were developed from researchers' observations and initial insights from the interview transcripts. Each interview was individually coded by two researchers to ensure coding consistency. Researchers met iteratively to discuss, refine, and group codes into themes until a coherent thematic framework was established. To ensure reliability and rigor, the four researchers compared codes across multiple rounds, resolving conflict through analytical discussions rather than statistical measures of coding accuracy, consistent with qualitative HCI research practices (McDonald & Forte, 2019). We recognize that our individual backgrounds may influence the outcomes of this research. We also acknowledge the potential for unequal tradeoffs between researchers, course facilitators, and the participants of this study, as they are from marginalized communities. Hence, researchers and the interviewers engaged in regular online meetings to collaboratively understand learners' responses and discuss updated findings from student interviews. During the thematic analysis process, we applied a qualitative research technique similar to member checking (Kullman & Chudyk, 2025) by consulting with interviewers and course facilitators from the same district as participants to validate and collaboratively discuss the findings.

## Findings

In this section, we illustrate the interactions of learners in the Hybrid (English and Leb-Lango) group, aligning with the mixed-methods analysis described above. Quotes from Hybrid learners are labeled with their respective language-use flags, "E" for English only, "L" for Leb-Lango only, and "X" for Explorers, along with their unique user ID. For transparency, we also report insights that may not be consistently reflected across all participants, but are included when they offered novel or significant contributions.

### Finding 1: The majority of learners initially chose to join the Hybrid group

When learners first registered for the remote course, a majority of students opted for the Hybrid condition of the course, with 43% (n = 1,270) of students selecting the Hybrid condition, 31% (n = 903) selecting English, and 26% (n = 758) selecting Leb-Lango. This initial Hybrid language preference can be explained by learners' perceptions of the importance of learning and knowledge in both languages for their livelihoods and futures. For instance, learners from the Hybrid group reported the importance of using English and Leb-Lango for communication in their daily lives. These learners reported engaging in constant negotiation when using both languages for communication with others, typically reserving English for communication outside their community and Leb-Lango for communication within their community. Several Hybrid students also explained that knowing both languages can enable learners to work and communicate with others. Many Hybrid learners also reported that having both languages available can enhance their comprehension of learning content. Of these students, several learners explained that their local language can enhance their understanding of learning content taught in English. For instance, E16 reported that "Sometimes, if you have used English and explained in your local language, you understand much better because you will get the clue and the content of what has been asked." E38 explained an advantage to flexibly transition from learning in English to Leb-Lango when explanations in English are difficult to understand. In contrast, E93 reported that not all learning can be explained in Leb-Lango, so having the option to learn in both languages can supplement these understanding gaps. Several Hybrid students also reported the value in comparing learning between English and Leb-Lango. For instance, X72 explained that they can relate terms for better understanding across both languages. Additionally, X04 reported that learning in both languages can accommodate students who are "literate or illiterate." Hence, the Hybrid learners demonstrate awareness of the benefits of learning with both English and Leb-Lango, which may have served as an initial motivator to join the Hybrid group of the course.

## Finding 2: In the course, the majority of Hybrid learners did not switch languages

Our subsequent analyses focused on specifically investigating the instructional language use of the 1,270 learners who chose the Hybrid option. In investigating the overall instructional language use of these learners, we found that 57.8% (n = 734) of Hybrid learners used English exclusively as an instructional language, 10.6% (n = 135) used only Leb-Lango, and 31.6% (n = 401) explored the use of both languages. Overall, the majority of Hybrid learners (68%) did not explore the use of both languages, with most learning with English exclusively. While finding 1 showed that the Hybrid learners recognized the value of both knowing and learning in both languages, they associated English with prestigious life opportunities and importance. This may indicate that the perceived benefits of learning in English outweighed the initial interest to learn with both instructional languages throughout the remote course. For instance, E08 reported that they chose to learn English to obtain a certificate for employment opportunities outside their community. Notably, a majority of learners who only learned with English are from higher educational experiences, specifically from secondary schooling levels, than students who only learned with Leb-Lango. Several Hybrid learners also expressed a preference for learning in English, citing past experiences of learning English in school. For example, E38 explained how they studied in a school "where people speak many languages, but the only language that was uniting people was English." E25 also showed that they attended school where their best friend does not speak Leb-Lango, so they communicate in English. E08 also reported that they chose to learn in English, as they attended secondary school up to an S4 level, where the main language of instruction is English. This learner explained that they did not want to learn in Leb-Lango as they associated Leb-Lango with people who lack schooling experience. These responses indicate a strong connection between English and education, where learners may show a preference to learn in English. In fact, this view is exemplified by E44, where they explain that they did not choose to learn in Leb-Lango because "English is compulsory".

Several Hybrid learners also reported difficulties navigating the use of both English and Leb-Lango in their remote course learning. For instance, L80 explained that it was difficult switching instructional languages: "[Switching] was not easy because if you only have a few chances, for example, in one or two steps, you can only choose one option. So you have to wait until you have been informed to switch or not, and then make the switch." E65 also reported, "I found it good to make a decision on using one language so that I could understand very well." When asked about changes in language preferences and how often learners switched languages throughout the course, several Hybrid learners simply stated that they only used one language to maintain their learning, which was primarily English. These responses may suggest that some Hybrid learners may have experienced cognitive overload when trying to engage with both English and Leb-Lango.

## Finding 3: Some Hybrid learners still explored using both languages in their learning

Overall, 31.6% of Hybrid learners explored the use of both English and Leb-Lango in the remote course (Explorers). However, the majority of these learners did not switch frequently between the two languages. 18.7% of learners who explored both languages changed languages at a given step. This suggests that while learners were willing to use both languages, most tended to consistently use the same language. Figure 2 (A) shows the proportion of Explorers who switched languages at each course step. Switching was highest at the introductory step (33.4%) and remained relatively high at Step 1 (33.1%). After Step 2, the proportion of learners switching languages decreased, reaching a low of 7.9% at Step 5. Small increases were observed at Step 4, Step 6, and Step 7, but overall, language switching was less frequent in the later steps of the course. These results suggest that learners may have experimented with language choice at the beginning of the course, possibly as they explored which language felt more comfortable or effective for comprehension. As learners progressed, most appeared to stabilize their choice and continued using a single language within each step.

Looking more closely at Explorers' language use trajectories, we found that their switching trajectories tended to be brief, with most students switching once or twice. We identified 10 distinct switching trajectories, capturing how learners moved between languages over time. The majority of trajectories occurred infrequently; therefore, we focus on the most common, which included (1) English → Leb-Lango → English, (2) English → Leb-Lango, (3) Leb-Lango → English → Leb-Lango, and (4) Leb-Lango → English. In Table 1, we report the proportions of language trajectories for Explorers. Overall, more learners began the course using English than Leb-Lango (63% vs. 37%) and ended the course using English than Leb-Lango (68% vs. 32%). Among learners who started in English, 62% continued using English at the end while 38% switched to Leb-Lango by the end. In contrast, most learners who began in Leb-Lango switched to English at the end (78%), with only 22% remaining in Leb-Lango. This indicates that most learners starting in English largely maintained their initial language, whereas those starting in Leb-Lango predominantly transitioned to English over the course.

**Table 1**
*Language trajectories: Start and End transitions among Explorers.*

| Start language | End language | No. of learners | Proportion (%) |
|---|---|---|---|
| English | Leb-Lango | 96 | 38 |
| English | English | 156 | 62 |
| Leb-Lango | Leb-Lango | 33 | 22 |
| Leb-Lango | English | 116 | 78 |

*Proportions are calculated relative to the total number of learners who started in the given language.*

While learners did not switch frequently, we found that exploring both languages even once was associated with longer course persistence (Figure 2 (B)). Although learner participation in the course persistence generally declined over the course, with early steps showing the highest proportion of students remaining and later steps showing lower persistence (all step coefficients significant, $p < .001$), Explorers persisted the most: 46% remained until the exam, compared to 41% of English-only learners ($b = -1.34$, $p < .001$), and 34% of Leb-Lango only learners ($b = -2.27$, $p < .001$).

**Figure 2**
*Proportion of learners categorized as Explorers who switched languages at each step of the course (A), and Proportion of Hybrid learners remaining at each course step by language behavior (B).*

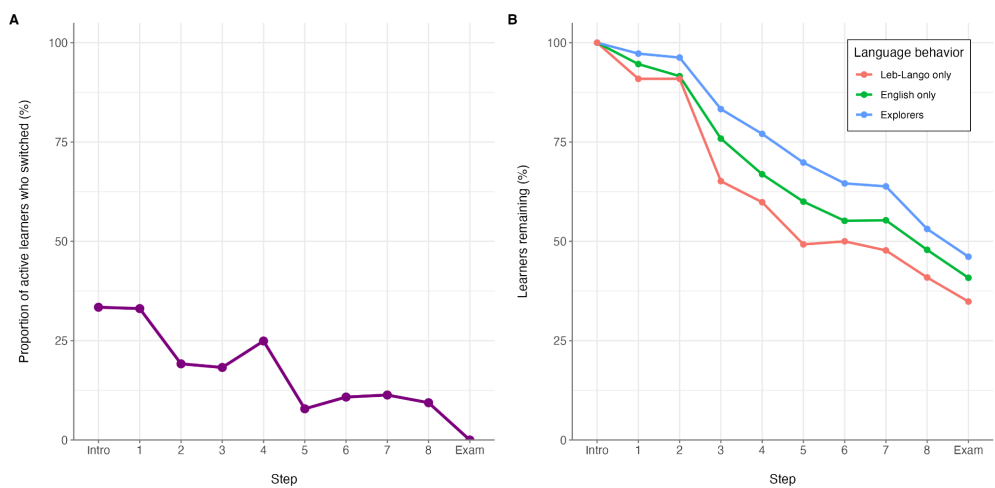

## Finding 4: Hybrid Learners had to consider important language complexities

Direct accounts from Hybrid learners revealed the complexities of navigating English and Leb-Lango, which we organized into three key interactions. These interactions also underscored the importance of using both languages in learners' remote course experience.

### Course Content and Terminology

When asked about moments when learners experienced changes in language preferences, several learners reported relying on Leb-Lango instruction when learning certain terms introduced in English became difficult. For instance, E52 reported that "There were some [instances] when the English was big, so I didn't understand it and changed to the local language." E97 also reported that "if I don't understand something in English, I wait for Leb-Lango lessons because I can understand better... There were some hard words." Though these Hybrid learners are indicated as English-only learners in their profile, they reported instances where they relied on the Leb-Lango radio broadcasts to understand terms that were challenging in English. Several Hybrid learners reported that when they were introduced to the concept of "criteria and constraints" that engineers consider when creating a product, they struggled to understand this concept in English. In fact, several Hybrid learners reported drawing on Leb-Lango instruction to provide explanations and guidance in creating their solar food dryers. When asked about facing difficulties learning in English, X72 explained that they preferred for practical activities to be explained in Leb-Lango. When asked at which parts of the course Leb-Lango was helpful, X63 reported that understanding how to "draw a plan" as part of Step 4 (the Plan Step) in Leb-Lango was useful. Conversely, Hybrid learners also reported difficulties learning certain terms in Leb-Lango, in which some terms are "forcefully translated from English" (L58). For example, X63 explained that "engineering" as a concept is more commonly used in English and has no direct Leb-Lango equivalent. E25 explained that though the concept of "technology" is "not very easy", it is "actually useful" when explained in English. Notably, E16 explained that they "understand English is much better, as our language, Leb-Lango, is somehow complicated, in which the vocabulary needs people who are old in it." This implies that though learners themselves primarily spoke Leb-Lango in their daily lives, efforts to maintain correctness in meaning resulted in Leb-lango terms that stayed close to their original form, which may not be used colloquially by younger learners as they may be more familiar with the English equivalents. Additionally, X73 provided a practical insight into why English provides better comprehension of certain terms and concepts, as "reading Leb-Lango words are long compared to English words." Hence, learners in the Hybrid group shared reciprocal perspectives on learning in English versus Leb-Lango. While learning in English adds a layer of complexity, learning in Leb-Lango also presents inherent challenges, as not all terms and concepts were clearly understood.

<u>Literacy and prior schooling</u>
The Hybrid learners also reported a significant challenge reading and writing in Leb-Lango. E47 reported that they chose to learn in English as "it is easy for me because with Leb-lango I take a while to understand it… I don't understand it that easily and even writing it is hard on my side." E44 explained that they encountered difficulties reading in Leb-Lango, further explaining that their literacy did not align with their learning. Difficulties accessing literacy in Leb-Lango can stem from two sources. Firstly, most pre-colonial societies in Sub-Saharan Africa had oral traditions, whose languages had no formal written form (Abdi, 2007). During the post-colonial period, many African countries, including Uganda, faced major language planning decisions, including the development of new writing systems for unwritten local languages (Nankindu, 2020). Secondly, local languages are provided as languages of instruction only within early primary education, whereas students mainly engage with English for the rest of their schooling. This view is evidenced by E75, who explained that they mainly used English instruction in the remote course "because reading in Leb-Lango is not easy though I know how to speak it, but reading is very hard… With English, I have studied in it, which is why I understand it very well." Literacy in Leb-Lango may still be relatively unfamiliar; under the country's current instructional policies, educational technology that includes written instruction may be more accessible in English.

<u>Leb-Lango and Community Use</u>
Several Hybrid learners also demonstrated careful negotiation in navigating Leb-Lango during their remote course experience, as a means of engaging with and contributing to their community. For instance, E38 explained that while they mainly learned in English, there was a moment when they tuned into the Leb-Lango radio broadcast for their grandmother to also listen to the course content. When asked about future language preferences in a similar course, X10 reported wanting to learn in Leb-Lango as they "always sit with the young kids near me during learning hours and they also want to listen to the program… the vocabulary the course uses, sometimes I fail to understand it [in English] and when the kids ask me I also fail to translate for them." These learners demonstrate the importance of Leb-Lango as a means to engage their family and community in their learning experiences. Learners also showed the importance of using their local language to channel their knowledge gained from the remote course back to their community. E38 explained the importance of using their local language to teach and communicate with village farmers and members who typically lack educational experiences. E08 also reported "It is also good for us to know the exact explanation. If you are to gather the community together, you have to explain to them how the machine can work because some people were asking me that they don't understand what the solar food dryer I am making is for. I have to explain to them in Leb-lango so that they can understand. If I just call it the solar dryer or solar food dryer, they can't understand."

## Finding 5: Hidden interactions showed several Hybrid learners using both languages
Several Hybrid learners revealed employing both English and Leb-Lango through different practices in their remote course experience. For instance, learners showed a preference for listening to the course content in Leb-Lango via radio broadcasts and interacting with the course content and responding to questions in English on their mobile phones: "When it comes to learning over the radio, I actually use Leb-Lango. But when I'm using the phone, I use English… Because it's much easier using English with the phone and listening over the radio when they're speaking in Leb-Lango, because I understand it much better" (E25). A similar approach is reported by E47 in which they answered questions and weekly tests in English on their phones and listened to lesson content in Leb-Lango: "I answer the daily tests and for the weekly, it is in English. However, I like listening to the one in Leb-lango at 2 PM." These learners adopted a bilingual approach, processing content more intuitively by listening to content in their local language, while using English to read and engage with content on their mobile phones, where literacy may be more accessible and content is more concise. Learners also reported listening to radio broadcasts in both English and Leb-Lango as they are delivered on the same day. E08 reported "waiting for the Leb-Lango time" to know that they "failed to understand during the English hour", especially when learning to make the first prototype. E71 followed a similar approach in using the Leb-Lango radio broadcasts to address gaps in understanding from the English broadcasts. E75 also reported consistently following both language broadcasts, but maintaining the use of one language on their phone. In contrast, L80 registered for each language on a separate SIM card to access course content in both languages. While these Hybrid learners are identified as single-language users in the log data, they demonstrated other alternative bilingual learning strategies to support their understanding of the learning material.

## Discussion and Conclusion

Our work unpacked learners' interactions as they navigated both colonial and local languages in their remote course experience. At a surface level, our preliminary findings indicate that a plurality of learners initially appreciated the option to learn in both English and Leb-Lango in the remote course. However, in our investigation analyzing the shifts in Hybrid learners' use of English and Leb-Lango throughout the remote course, our subsequent findings show that most learners demonstrated a tendency to maintain the use of one language, primarily English. Aligning with prior work and the continual preference for colonial languages of instruction, many learners additionally emphasized the association between English and prestige, indicating that the perceived benefits of learning in English may have changed the appeal of learning in both languages.

This preference for English might tempt developers to only use English in EdTech programs for learners in similar remote contexts. Our data does not suggest that learners should exclusively learn in English, however, as a significant portion of Hybrid learners (31%) did explore instruction in both English and Leb-Lango throughout the remote course. In fact, a more in-depth analysis revealed that these learners showed higher course persistence than Hybrid students who only learned in English or Leb-Lango. While this does not imply causality, the higher persistence may be due to Explorers having a better understanding of which language is helpful for their learning. This is likely because they directly experienced the course in both languages at the beginning of the course. In contrast, learners who only chose to experience learning in one language and may have missed out on potential benefits of learning in the other language, in which encountering difficulties and limitations learning in one language may be associated with lower course persistence. Furthermore, learners' experiences with both languages indicated the importance of using nuanced ways to navigate their local language and English, learners sharing complexities with terminology and content in English and Leb-Lango, difficulties with literacy in Leb-Lango, and engaging their learning with their community in Leb-Lango.

Hence, our work highlights the importance of EdTech and its ability to support the complementary use of both colonial and local languages to make education more accessible and meaningful for learners within marginalized multilingual regions. While our findings suggest that colonial languages may serve a practical instructional role in EdTech, we recommend that EdTech developers avoid marginalizing local language practices and instead ensure that local languages are actively supported within EdTech environments. Developers should raise awareness among remote learners of the benefits of learning in their local language, as our study shows that multilingual learning was associated with higher course persistence. We recommend that EdTech developers deliver instruction and educational content available in both local and colonial languages, giving learners agency to flexibly navigate learning in both languages in ways that feel most comfortable to them, a view that is also corroborated by the observed positive learning benefits shown by prior studies on translanguaging practices (Nyimbili & Mwanza, 2020) and multilingual mobile learning tools (Jantjies & Joy, 2012; Jantjies & Joy, 2013). Building on this foundation, prior literature has also underscored the need to better understand how multilingual instruction should be meaningfully integrated in EdTech. To address this gap, our study offers a more concrete approach, recommending that learners choose a primary language of instruction while leveraging an additional language of support when needed. This approach respects learners' language preferences, offers translanguaging support, and provides multilingual instruction at scale through EdTech capabilities. For instance, with this approach, the appeal of learning in colonial languages can be leveraged to engage learners and support instruction, while learners also have access to their local language for more learning support. Additionally, our findings provide initial insight into how EdTech modalities can support multilingual instruction, as several Hybrid learners reported a preference for listening to radio broadcasts in their local language and for interacting with and reading content on their phones in English. This encourages further investigation into how to prioritize instructional languages across different modalities of EdTech to meet the linguistic needs of multilingual learners.

## References

Abdi, A. A. (2007). Oral Societies and Colonial Experiences: Sub-Saharan Africa and the De Facto Power of the Written Word. International Education, 37(1).

Altinyelken, H. K. (2010). Curriculum change in Uganda: Teacher perspectives on the new thematic curriculum. International journal of educational development, 30(2), 151-161.

Breazeal, C., Morris, R., Gottwald, S., Galyean, T., & Wolf, M. (2016, April). Mobile devices for early literacy intervention and research with global reach. In Proceedings of the third (2016) ACM conference on learning@ scale (pp. 11-20).

Braun, V., & Clarke, V. (2006). Using thematic analysis in psychology. Qualitative research in psychology, 3(2), 77-101.

Buhl-Wiggers, J., Kerwin, J. T., de la Piedra, R. M., Smith, J., & Thornton, R. (2023, May). Reading for life: Lasting impacts of a literacy intervention in Uganda.

Criper, C. and Ladefoged, P. (1971) Linguistic complexity in Uganda. In W.H. Whiteley (ed.) Language Use and Social Change: Problems of Multilingualism with Special Reference to Eastern Africa (145–59). Oxford: Oxford University Press.

Jantjies, M., & Joy, M. (2012). Multilingual mobile learning: a case study of four South African high schools. In Proceedings of the 11th International Conference on Mobile and Contextual Learning 2012 (pp. 208-211). CEUR-WS.

Jantjies, M., & Joy, M. (2013). Mobile learning through indigenous languages: learning through a constructivist approach. QScience Proceedings, 2013(3), 14.

Khan, S., Hwang, G. J., Azeem Abbas, M., & Rehman, A. (2019). Mitigating the urban–rural educational gap in developing countries through mobile technology‑supported learning. British Journal of Educational Technology, 50(2), 735-749.

Kukulska-Hulme, A., Giri, R. A., Dawadi, S., Devkota, K. R., & Gaved, M. (2023). Languages and technologies in education at school and outside of school: Perspectives from young people in low-resource countries in Africa and Asia. Frontiers in Communication, 8, 1081155.

Kullman, S. M., & Chudyk, A. M. (2025). Participatory member checking: A novel approach for engaging participants in co-creating qualitative findings. International Journal of Qualitative Methods, 24, 16094069251321211.

Kwon, C., Butler, R., Stamper, J., Ogan, A., Forcier, A., Fitzgerald, E., & Wambuzi, S. (2023). Learning analytics for last mile students in Africa. In Proceedings of the 13th Learning Analytics and Knowledge Conference.

Liddicoat, A. J., & Heugh, K. (2014). Educational equity for linguistically marginalised students. In The Routledge handbook of educational linguistics (pp. 79-91). Routledge.

McDonald, N., Schoenebeck, S., & Forte, A. (2019). Reliability and inter-rater reliability in qualitative research: Norms and guidelines for CSCW and HCI practice. Proceedings of the ACM on human-computer interaction, 3(CSCW), 1-23.

Mohammed, P., & Mohan, P. (2013). A case study of the localization of an intelligent tutoring system. In Proc. 2nd International Workshop on Learning Technologies for the Developing World in conjunction with AIED 2013, Memphis, USA, July 9–13.

Mohammed, P., & Mohan, P. (2015). Dynamic cultural contextualisation of educational content in intelligent learning environments using ICON. International Journal of Artificial Intelligence in Education, 25(2), 249-270.

Msimanga, A., & Lelliott, A. (2013). Talking Science in Multilingual Contexts in South Africa: Possibilities and challenges for engagement in learners home languages in high school classrooms. International Journal of Science Education, 36(7), 1159-1183.

Muthwii, M. J. (2004). Language of instruction: A qualitative analysis of the perceptions of parents, pupils and teachers among the Kalenjin in Kenya. Language Culture and Curriculum, 17(1), 15-32.

Nankindu, P. (2020). The history of educational language policies in Uganda: Lessons from the past. American Journal of Educational Research, 8(9), 643-652.

National Curriculum Development Centre. (2006). The National Primary School Curriculum for Uganda, Primary 1.

Ndebele, H. (2014). Promoting indigenous African languages through information and communication technology localisation: A language management approach. Alternation Special Issue, 13, 102-127.

Nkoala, S. (2024). Educators' Experiences of using multilingual pedagogies during emergency remote teaching: a case study of South African universities. International Journal of Multilingualism, 21(1), 534-547.

Nomlomo, V. S. (2007). Science teaching and learning through the medium of English and isiXhosa: A comparative study in two primary schools in the Western Cape (Doctoral dissertation, University of the Western Cape).

Nyimbili, F., & Mwanza, D. S. (2020). Quantitative and qualitative benefits of translanguaging pedagogic practice among first graders in multilingual classrooms of Lundazi district in Zambia.

Omojola, O. (2009). English-oriented ICTs and ethnic language survival strategies in Africa. Global Media Journal-African Edition, 3(1), 33-45.

Piper, B., Zuilkowski, S. S., & Ong'ele, S. (2016). Implementing mother tongue instruction in the real world: Results from a medium-scale randomized controlled trial in Kenya. Comparative Education Review, 60(4), 776-807.

Schmied, J., & Mesthrie, R. (2008). East African English (Kenya, Uganda, Tanzania): Phonology. Varieties of English, 4, 150-163.

Skutnabb‑Kangas, T., & McCarty, T. L. (2008). Key concepts in bilingual education: Ideological, historical, epistemological, and empirical foundations. In Encyclopedia of language and education (pp. 1466-1482). Springer, Boston, MA.

Trudell, B. (2007). Local community perspectives and language of education in sub-Saharan African communities. International Journal of Educational Development, 27(5), 552-563.

Uganda Bureau of Statistics. (2024). The national population and housing census 2024: Final report, volume 1 (Main). https://www.ubos.org/wp-content/uploads/2024/12/National-Population-and-Housing-Census-2024-Final-Report-Volume-1-Main.pdf

Verlumun Celestine, G., Talabi, F. O., Talabi, J. M., Adefemi, V. O., Bello, S. A., Oluyemi, A. A., & Destiny Apuke, O. (2024). Learning through interaction: impact of interactive radio instructions in improving literacy skills of out-of-school children in IDP camps in Nigeria. Interactive Learning Environments, 32(3), 1058-1067.

Wahyuni, E., Khan, O., & Raza, A. (2024). The Effectiveness of Mobile Learning in Remote and Rural Areas. Journal International Inspire Education Technology, 3(3), 253-264.

Wildsmith-Cromarty, R., Dyer, C., & Modipa, T. (2023). Enhancing visibility of local African languages in South Africa through learning to read. Journal of Multilingual and Multicultural Development, 44(9), 860-876.

Wolff, H. E. (2017). Language ideologies and the politics of language in post-colonial Africa. Stellenbosch Papers in Linguistics Plus, 51(1), 1-22.

## Acknowledgements

We would like to thank Jacobs Foundation (Award 2023151900) for supporting this research. We would also like to thank our collaborator, Yiya Solutions, Inc., who provided the data, supplementary materials of the course, and extensive support for this work.